\documentclass[aip,graphicx, reprint]{revtex4-1}
\usepackage{graphicx}
\usepackage{bm}
\usepackage{amssymb}
\usepackage{amsmath}
\usepackage{natbib}
\usepackage{color}
\usepackage{hyperref}
\hypersetup{
  colorlinks=true,
  citecolor=blue,
  linkcolor=blue,
  urlcolor=blue}
\usepackage{textcomp} 

\draft 

\begin{document}


\title{High-resolution gamma-ray spectroscopy with a microwave-multiplexed transition-edge sensor array} 



\author{Omid Noroozian}
\email[Electronic mail: ]{omid.noroozian@nist.gov}
\affiliation{National Institute of Standards and Technology, Boulder, CO 80305, USA}
\affiliation{Center for Astrophysics and Space Astronomy, University of Colorado, Boulder, CO 80309, USA}

\author{John A. B. Mates}
\affiliation{National Institute of Standards and Technology, Boulder, CO 80305, USA}
\author{Douglas A. Bennett}
\affiliation{National Institute of Standards and Technology, Boulder, CO 80305, USA}
\author{Justus A. Brevik}
\affiliation{National Institute of Standards and Technology, Boulder, CO 80305, USA}
\author{Joseph W. Fowler}
\affiliation{National Institute of Standards and Technology, Boulder, CO 80305, USA}
\author{Jiansong Gao}
\affiliation{National Institute of Standards and Technology, Boulder, CO 80305, USA}
\author{Gene C. Hilton}
\affiliation{National Institute of Standards and Technology, Boulder, CO 80305, USA}
\author{Robert D. Horansky}
\affiliation{National Institute of Standards and Technology, Boulder, CO 80305, USA}
\author{Kent D. Irwin}
\affiliation{National Institute of Standards and Technology, Boulder, CO 80305, USA}
\author{Zhao Kang}
\affiliation{Department of Physics, University of Colorado, Boulder, CO 80309, USA}
\author{Daniel R. Schmidt}
\affiliation{National Institute of Standards and Technology, Boulder, CO 80305, USA}
\author{Leila R. Vale}
\affiliation{National Institute of Standards and Technology, Boulder, CO 80305, USA}
\author{Joel N. Ullom}
\affiliation{National Institute of Standards and Technology, Boulder, CO 80305, USA}


\date{\today}

\begin{abstract}
We demonstrate very high resolution photon spectroscopy with a microwave-multiplexed two-pixel transition-edge sensor (TES) array. We measured a $^{153}$Gd photon source and achieved an energy resolution of 63 eV full-width-at-half-maximum at 97 keV and an equivalent readout system noise of 86 pA/$\sqrt{\text{Hz}}$ at the TES. The readout circuit consists of superconducting microwave resonators coupled to radio-frequency superconducting-quantum-interference-devices (SQUID) and transduces changes in input current to changes in phase of a microwave signal. We use flux-ramp modulation to linearize the response and evade low-frequency noise. This demonstration establishes one path for the readout of cryogenic X-ray and gamma-ray sensor arrays with more than 10$^{3}$ elements and spectral resolving powers $R=\lambda/\Delta\lambda > 10^{3}$.
\end{abstract}


\maketitle 


Multiplexed readout of sub-Kelvin cryogenic detectors is an essential requirement for large focal plane arrays. Next-generation instruments for the detection of electromagnetic radiation from gamma-ray to far-infrared wavelengths will have pixel counts in the $10^3$--$10^6$ range and require readout techniques that do not compromise their sensitivity. To date, many instruments have used time-, frequency-, or code-domain SQUID multiplexing schemes \cite{Doriese_TDM_APL_2004, Cunningham_FDM_APL_2002, Stiehl_CDM_APL_2012}. One such instrument, the TES bolometer camera SCUBA2, has achieved background-limited sensitivity in $10^4$ pixels using time-domain multiplexing (TDM) \cite{Holland_SCUBA-2_MNRAS_2013}. Similarly, calorimetric gamma-ray/X-ray spectrometers that use TDM have reached excellent energy resolutions of $\delta E\approx 50$ eV at 100 keV in a 256-pixel array\cite{Bennett_microcal_SciRev_2012}. However, the scalability of these readout approaches is limited by the finite measurement bandwidth ($\sim10$ MHz) achievable in a flux-locked loop.

Kinetic Inductance Detectors (KIDs) \cite{Day_MKIDs_Nature_2003, Zmuidzinas_ResonatorsReview_ARCMP_2012}, on the other hand, provide a possible path to higher multiplexing factors. These devices are naturally frequency-multiplexed and the ultimate limit on the available bandwidth is many gigahertz, which is set by the readout cryogenic amplifier. Present limits in room-temperature electronics impose a 550 MHz bandwidth limit \cite{Mchugh_Readoutelecs_RSI_2012}, but this figure will improve steadily. However, the sensing element is part of a thin-film superconducting resonator, so readout and signal generation can be difficult to simultaneously optimize. This challenge is particularly severe for spectroscopic X-ray and gamma-ray detectors, which must stop high-energy photons and where spatial variation in the device response must be smaller than 0.1\%. X-ray and gamma-ray spectroscopy results achieved to date with KIDs are not yet compellingly better than conventional semiconducting detectors \cite{Mazin_X-ray_MKIDs_APL_2006, Moore_PhononMKID_APL_2012}.

Microwave SQUID multiplexing \cite{Irwin_MicrowaveSQUIDMux_APL_2004, Mates_MicrowaveSQUIDMux_APL_2008} (\textmu Mux) is a readout technique that potentially combines the proven sensitivity of TESs and the scalable multiplexing power found in KIDs. Microwave SQUID multiplexing uses radio-frequency (rf) SQUIDs coupled to high quality-factor ($Q$) microwave resonators and has sufficiently low noise to read out the most sensitive cryogenic detectors. Additionally, it allows independent optimization of the detector and the multiplexer, and provides signal modulation that evades low-frequency resonator noise. Previously, we demonstrated device-noise limited \textmu Mux readout of two 150-GHz TES polarimeter bolometers \cite{Mates_Thesis}. Here we demonstrate the use of \textmu Mux with TES gamma-ray detectors that respond to single incident photons. Specifically, we demonstrate spectroscopy with resolving powers $R \gtrsim 1500$ using \textmu Mux and a two-pixel TES array. Our achieved resolving power is close to an order of magnitude higher than state-of-the-art high-purity germanium (HPGe) detectors. Furthermore, the number of pixels can easily be scaled to values $> 10^3$ in future instruments to provide useful system-level count rates and collecting areas.

\begin{figure}[t]
\includegraphics[width=0.5\textwidth]{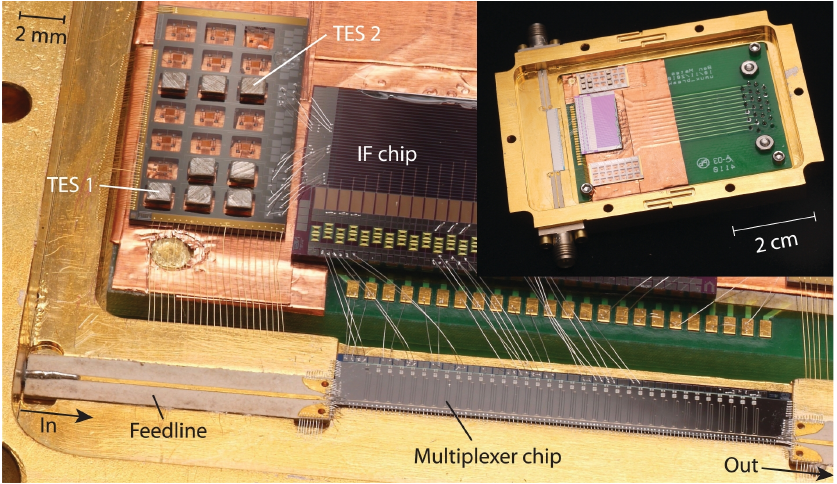}
\caption{Photograph of the microwave SQUID multiplexer and gamma-ray TESs with bulk Sn absorbers. TESs numbered 1 and 2 are wirebonded to SQUID input coils on the multiplexer chip through a central interface (IF) chip.  Up to 35 resonators on the \textmu Mux chip are read out using the single microwave feedline at bottom. Inset shows the full device box with two microwave SMA connectors.} \label{figure:uMux_photo}
\end{figure}

The main goal of this experiment was to demonstrate the readout of two gamma-ray TESs using microwave SQUID multiplexing with little degradation in energy resolution and in a scalable fashion that establishes a path for instruments with much larger detector count. A photograph of the device used is shown in Fig. \ref{figure:uMux_photo}. The sensor chip consists of 21 TES microcalorimeters similar in design to devices reported by \textit{Bennett et al.}\cite{Bennett_microcal_SciRev_2012}, but modified for higher energy resolution and smaller dynamic range. The TESs are made from a Mo-Cu bilayer with $T_{\text{c}}=107$ mK and heat capacity $C_1\approx 2.3$ pJ/K, and are placed on a $1.38 \times 1.38\ \text{mm}^2$ Si$_3$N$_4$ membrane 1 \textmu m thick that provides a thermal conductance $G_1\approx 2.2$ nW/K. Bulk absorbers of polycrystaline Sn with dimensions of $1.1\times 1.1\ \text{mm}^2 \times 250$ \textmu m are glued using Stycast 1266 to SU8 epoxy posts on the SiN membrane.  The thermal conductance between absorber and TES was approximately  $G_2\approx 31$ nW/K. The absorber has a heat capacity of $C_2\approx 6$ pJ/K and provides $\sim 27\ \%$ absorption efficiency for a 100 keV photon.

An interface (IF) chip was used to provide a bias shunt resistance of $R_{\text{sh}}=0.33$ m$\Omega$ in parallel with each TES and a wirebond-selectable Nyquist inductor, $L_{\text{N}}$, in series to increase the pulse rise-times. The value for $L_{\text{N}}$ can be selected from values of 0, 270, and 690 nH. Although the IF chip was not specifically designed for this experiment it provided reasonable values of resistance and inductance. We used $L_{\text{N}} = 690$ nH for TES 1 and $L_{\text{N}} = 0$ nH for TES 2.

The multiplexer chip consists of 35 quarter-wave coplanar waveguide (CPW) microwave resonators made from a 200 nm thick Nb film deposited on high-resistivity silicon with $\rho > 10\ \text{k}\Omega\ \text{cm}$. The resonator CPWs have a 10 \textmu m center strip and 6 \textmu m gap widths. Adjacent resonances are separated by $\sim6$ MHz and are centered around 5.5 GHz with coupling quality factors $11,500 \lesssim Q_{\text{c}} \lesssim 34,000$ and internal quality factors $48,000 \lesssim Q_{\text{i}} \lesssim 120,000$. We chose two resonances at 5.503 GHz and 5.566 GHz with $Q_{\text{c}}$'s of 12,000 and 32,000 and $Q_{\text{i}}$'s of 80,000 and 110,000 to read out TES 1 and 2 respectively. The short circuit end of each resonator inductively couples to an rf SQUID (with geometric inductance $L_{\text{s}} \approx 20$ pH and critical current $I_{\text{c}} \approx 5$ \textmu A) that acts as a flux-dependent nonlinear inductor. The rf SQUID transduces a change in input flux into a change of resonance frequency. In turn, the TES current couples flux into the SQUID through an input coil with mutual inductance $M \approx 88$ pH. A common flux line is inductively coupled to all the SQUIDs to provide flux-ramp modulation \cite{Mates_FluxrampModulation_JLTP_2012} ability (see below). The details of the chip can be found in Mates's PhD thesis \cite{Mates_Thesis}.

The chips were mounted in a gold-plated copper sample box. A G-10 circuit board provides DC connectivity for the chips. Two Duroid circuit boards with microstrip to CPW transitions connect the microwave input and output lines to the CPW feedline on the resonator chip and to SMA connectors on the box. In this unoptimized setup long aluminum wirebonds ($\lesssim 1$ cm) connect the SQUID input coils to the IF chip and the TESs, and bring in the DC bias; these long free-space connections are a likely source of 1/f noise. Gold wirebonds were used for heat-sinking the TES chip. A small hole in the copper box lid (not shown here) above the TES chip increases the gamma-ray flux reaching the absorbers.

The sample box was mounted inside a cryostat and was connected to a pulse-tube backed adiabatic demagnetization refrigerator (ADR), and its temperature was regulated at 85 mK using the ADR magnet. During the initial cooldown from 300 K a magnetic shield made from mu-metal was placed around the cryostat to avoid trapping earth's field inside the resonators, SQUIDs, and Sn absorbers. This shield was removed after reaching base temperature to increase the gamma-ray count rate. Once this shield was removed, there was no magnetic shielding for the experiment. The IV curves for some of the TESs showed distortions from magnetic field trapped in the nearby Sn absorber, which resulted in lower pulse heights. This flux trapping likely occurred when the ADR was cycled to reach 85 mK. However, TESs 1 and 2 showed good IV characteristics with no evidence of flux trapping. Successful unshielded operation at 85 mK (after removing the mu-metal) in the presence of earth's field and the ADR field bodes well for future robustness. The final gamma-ray path contained a 0.8 mm thick carbon fiber window in the cryostat vacuum shell and three access windows cut into the 60 K and 3 K radiation shields and the Cu box lid. All three access windows were covered with 0.1 mm thick aluminum tape. We positioned a weak $^{\text{153}}$Gd radioisotope source outside the carbon fiber window approximately 10 cm from the detectors to provide a photon count rate of $\sim 0.75$ Hz per detector.

\begin{figure}[t]
\includegraphics[width=0.5\textwidth]{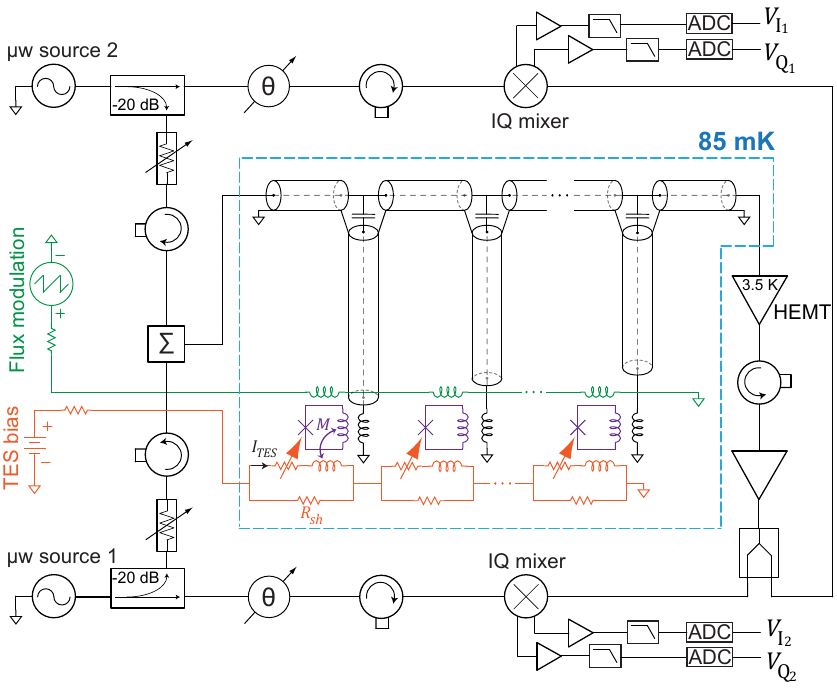}
\caption{Circuit schematic for multiplexed gamma-ray spectroscopy. Resonator circuitry is shown in black, rf-SQUID components in purple,  TES components in orange, and flux ramp components in green. Dashed blue line contains circuit components at the cold stage of the ADR. Microwave attenuators at various stages have been omitted for clarity.} \label{figure:uMux_circuit}
\end{figure}

\begin{figure}[t]
\includegraphics[width=0.5\textwidth]{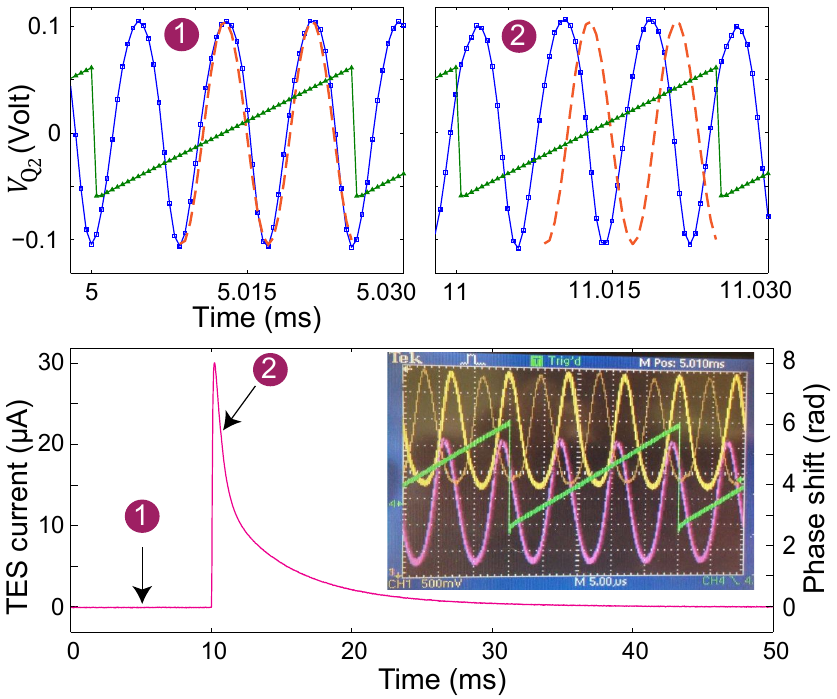}
\caption{The two top figures show the measured quadrature-phase component $V_{\text{Q}_2}$ (blue circles) of the microwave signal, the flux-ramp modulation sawtooth signal $V_{\text{fr}}$ (green triangles), and a pure sinusoidal waveform (orange dashed) over a 32 \textmu s window sometime before (1) and during (2) a 97 keV gamma-ray pulse in TES 2. The bottom panel shows the corresponding TES current (left axis) and resonance phase shift (right axis) pulse after demodulation. The inset is a snapshot from an oscilloscope simultaneously measuring $V_{\text{Q}_1}$ for TES 1 (top yellow) and $V_{\text{Q}_2}$ for TES 2 (bottom purple) just before digitization (see Fig. \ref{figure:uMux_circuit}). The faint phase-shifted yellow waveform represents a gamma-ray event. A real-time movie showing gamma-ray events in both TESs is available online \cite{uMux_Movie}.}  \label{figure:uMux_modulation}
\end{figure}
\begin{figure}[!t]
\includegraphics[width=0.5\textwidth]{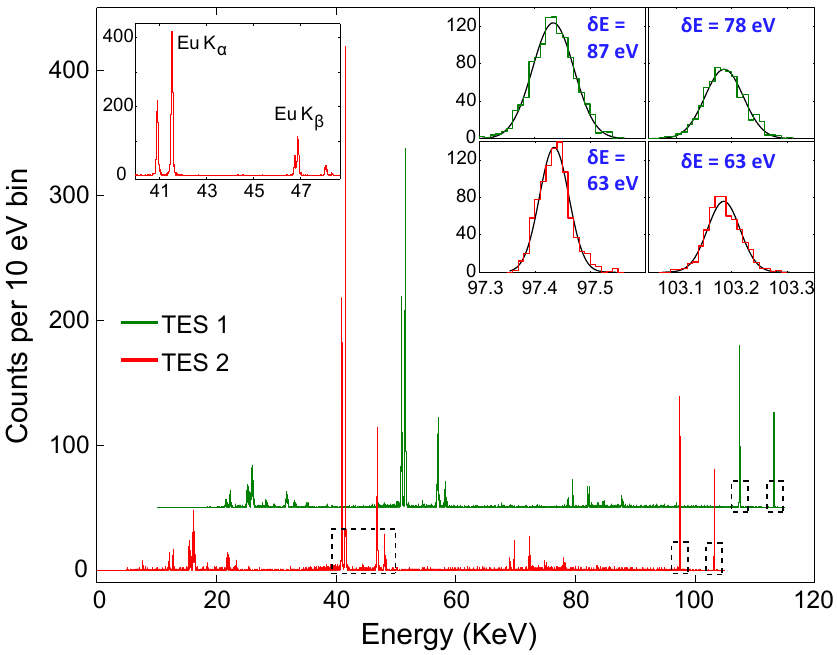}
\caption{Energy spectrum of a $^{\text{153}}$Gd source measured using our two-pixel TES array and the microwave SQUID multiplexer. The spectrum from the TES 1 pixel has been horizontally and vertically shifted for clarity. The top right four insets show a zoom-in of the two 97.4 keV and 103.2 keV photopeaks and corresponding Gaussian fits for TES 1 (top green) and TES 2 (bottom red). The full-width-at-half-maximum (FWHM) energy resolution $\delta E$ for each peak is indicated. The top left inset shows a zoom-in from 40--49 keV where europium K$_\alpha$ and K$_\beta$ complexes can be seen.} \label{figure:uMux_spectrum}
\end{figure}

The circuit diagram for simultaneous readout of two TESs is shown in Fig. \ref{figure:uMux_circuit}. The number of sensors in our demonstration was set by the availability of room-temperature microwave electronics but, as shown in Fig. \ref{figure:uMux_circuit}, the circuit architecture is compatible with a much larger number of sensors that share the same microwave feedline, and can easily be scaled by using software-defined radio electronics \cite{Duan_Opensource_readout_SPIE_2010, Mchugh_Readoutelecs_RSI_2012}. In our experiment two microwave signal generators tuned close to the frequencies of the resonators inject two tones into the feedline. Each tone's frequency and power is adjusted to optimize the signal-to-noise \cite{Mates_Thesis}. The final readout powers were $P_{\text{\textmu w}}=-69$ dBm and -73 dBm at the \textmu Mux chip feedline for the resonators connected to TESs 1 and 2, respectively. The TESs are voltage biased in their resistive transition at $\sim20\ \%$ of their normal-state resistance $R_\text{n}$ by use of a single DC bias signal. A gamma-ray photon event in a TES increases the temperature and therefore the resistance of the TES, which in turn reduces the current $I_{\text{TES}}$ passing through the TES. This time-dependent current applies flux $\Phi$ in the SQUID loop as $\Phi = M I_{\text{TES}}$. The value of $M$ sets the transduction gain. The SQUID inductance is periodic with flux, as $L\approx L_{\text{s}} + L_{\text{J}}/\cos\phi$ where $L_{\text{J}}=\Phi_0 / (2\pi I_{\text{c}})$ is the Josephson inductance, $\phi=2\pi \Phi/\Phi_0$ is the phase shift across the junction, and $\Phi_0$ is the magnetic flux quantum. Therefore, a large input flux signal can cause an excursion of several flux periods in the SQUID inductance and consequently in the resonance frequency. Changes in the resonance frequencies of the two resonators change the complex microwave transmission $S_{21}$ across the feedline. These changes are amplified with a cryogenic high-electron-mobility-transistor (HEMT) amplifier and further amplified at room temperature. Two IQ mixers then downconvert the signals in each channel using copies of the original microwave tones. The in-phase ($V_{\text{I}}$) and quadrature-phase ($V_{\text{Q}}$) signals are then digitized at a sample rate of 2 MHz in a computer. We used only the $V_{\text{Q}}$ signal components, rotating the resonance IQ planes with phase shifters such that they coincide with the electronics IQ plane in each channel.

In order to linearize the signal response we implemented flux-ramp modulation \cite{Mates_FluxrampModulation_JLTP_2012} by applying a $f_{\text{s}} = 40$ kHz sawtooth flux-ramp signal $V_{\text{fr}}$ (see Fig. \ref{figure:uMux_modulation}) to all of the SQUIDs. The amplitude of the ramp is tuned such that it provides $\sim3\Phi_0$ of flux per ramp period and modulates $V_{\text{Q}_1}$ and $V_{\text{Q}_2}$ at a carrier frequency of $f_{\text{c}}\approx120$ kHz. Since $f_{\text{c}}$ is significantly larger than the frequency content of a gamma-ray pulse, the phase shift $\phi$ of $V_{\text{Q}}$ during each ramp period is effectively constant and proportional to the current signal as $I_{\text{TES}} = \Phi_0\phi/(2\pi M)$. The Nyquist signal sampling rate is therefore 40 kHz. Flux-ramp modulation has the added benefit that the signal is upconverted to frequencies above the low-frequency two-level system (TLS) noise that is intrinsic to the resonator \cite{Gao_NoisePropertiesOfResonators_APL_2007, Barends_dielectricNoiseNbTiN_APL_2008, Noroozian_TLS_LTD13_2009, Zmuidzinas_ResonatorsReview_ARCMP_2012}. The two top panels in Fig. \ref{figure:uMux_modulation} show the measured $V_{\text{Q}}(t)$ component for the TES 2 channel during a 32 \textmu s window before and after a gamma-ray pulse (blue circles). In order to demodulate the data stream we apply on-the-fly Fourier analysis to extract the phase shift $\phi$ of the fundamental frequency ($f_{\text{c}}$) component in $V_{\text{Q}}$ for each 25 \textmu s ramp window as $\phi = \arctan{\big( \frac{\sum{V_{\text{Q}}(t) \sin{2\pi f_{\text{c}} t}}} {\sum{V_{\text{Q}}(t) \cos{2\pi f_{\text{c}} t}}}\big) }$. The sine and cosine are defined over an integer number of periods (2 in this case) ending at the ramp reset (see orange dashed line in Fig. \ref{figure:uMux_modulation} top panels). The remaining $\sim 1$ period of $V_{\text{Q}}$ at the start of the ramp is ignored to prevent the unwanted transient behavior caused by the ramp reset from contaminating the data. $f_{\text{c}}$ is measured from a sinusoidal fit to the flux-ramp response when the TES is superconducting (i.e., zero input flux signal). The phase shift and corresponding TES current for a measured 97 keV pulse after demodulation are shown in the bottom panel in Fig. \ref{figure:uMux_modulation}.

\begin{figure}[t]
\includegraphics[width=0.5\textwidth]{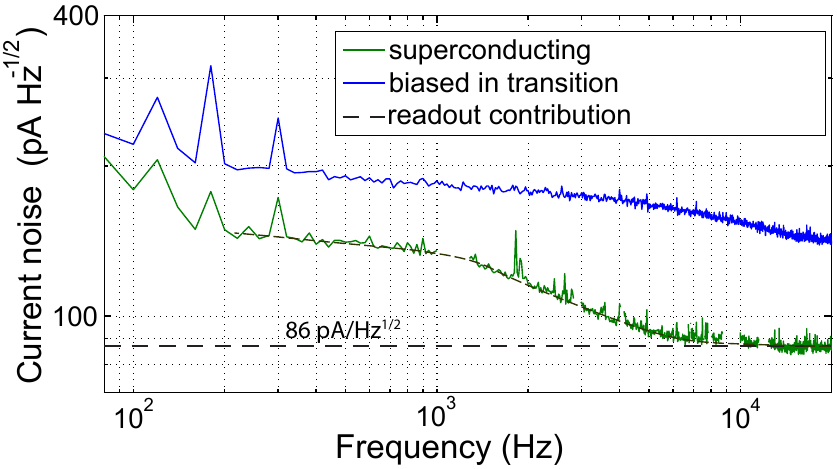}
\caption{Measured noise from the microwave multiplexer and TES device for channel 2. The plot is given in terms of the equivalent current noise referred to the TES. The blue line shows the total system noise when the TES is biased in the transition at 20 $\%$ of $R_\text{n}$. The green line shows the total system noise when the TES is superconducting. Far above the L/R roll-off the $\sim86$ pA/$\sqrt{\text{Hz}}$ remaining noise is from the \textmu Mux readout system. Spikes due to noise pickup have been removed from data for clarity.} \label{figure:uMux_noise}
\end{figure}

After simultaneously collecting data for TES 1 and 2 over a seven-hour period we excluded pulse records contaminated by pile-up and other nonidealities. An optimal filter was then applied determined by the power spectral density (PSD) of measured noise (below; see Fig. \ref{figure:uMux_noise}) and average pulse shape \cite{Szymkowiak_OptimalFilter_JLTP_1993}. A correction was then made due to drift in the peak energies over the time of the measurement. Finally, we calibrated the energy scale using four known spectral features. The resulting spectra for two simultaneously measured TESs are shown in Fig. \ref{figure:uMux_spectrum} where the two most prominent $^{153}$Gd gamma-ray photopeaks at 97.4 keV and 103.2 keV are shown in the upper right insets. Weighted Gaussian fits to these lines give FWHM energy resolutions of $\delta E = 63.0\pm2.2$ eV and $63.8\pm 2.9$ eV for TES 2, and $\delta E=87.3\pm 2.6$ eV and $78.1\pm 3.6$ eV for TES 1, for the 97.4 and 103.2 keV peaks, respectively. These resolutions are close to the expected resolution of 55 eV for TES 2 and 66 eV for TES 1, which were obtained from the noise PSD and average 97 keV pulse shapes \cite{Moseley_xrayThermalDetectors_JAP_1984}. The difference can be attributed to several factors, including residual pulse-tube noise at lower frequencies, uncorrected gain drift, and position-dependence of the TES response, all of which degrade the performance. More specifically, the pulse-tube noise was non-stationary such that consecutive gamma-ray pulses did not always experience the same noise level. The upper left inset shows the Eu $K_\alpha$ and $K_\beta$ complexes. The remaining lines include Sn X-ray escape peaks and fluorescence from the gold plating of the sample box.

In order to evaluate the noise performance of our readout circuit we made noise measurements with TES 2 biased in the transition at 20 $\%$ of $R_\text{n}$ and also unbiased (superconducting), as shown in Fig. \ref{figure:uMux_noise}. When biased, the TES noise contribution rolls off at $\sim 20$ kHz. When the TES is superconducting the shunt resistance contribution to the total noise rolls off at $\sim 3$ kHz, above which the remaining $\sim 86$ pA/$\sqrt{\text{Hz}}$ noise is the contribution from the readout circuit. The electrical roll-offs are consistent with the shunt and TES resistance and inductance values (including approximate wirebond inductance) in the circuit. It can be inferred that the difference in noise power between the blue curve and the readout noise level obtained by subtracting in quadrature is the TES contribution to the noise. To confirm this, we compared the inferred low-frequency TES noise level to an independent measurement of TES noise in a similar device performed using a TDM readout. The TDM data give a TES noise level of 150 pA/$\sqrt{\text{Hz}}$, which closely matches the inferred TES noise level of 140 pA/$\sqrt{\text{Hz}}$. From this we conclude that the \textmu Mux readout noise is a factor of 1.6 below the signal-band TES noise. We also conducted a set of TDM measurements where the readout noise was effectively negligible and where, consequently, we expected slightly better energy resolution. Indeed, measurements using TDM with 12 TESs from the same fabrication batch achieved resolutions of $\sim 50 \pm 8$ eV. The slight resolution advantage of the TDM results is consistent with the quadrature noise penalty from the current implementation of \textmu Mux readout. Although our achieved resolution of 63 eV is already sufficient for most spectroscopic applications, we expect that a number of simple modifications to the readout circuit should provide even better performance. Lower readout noise comfortably below the TES noise can be achieved with a larger input coil mutual inductance $M$ that boosts the TES signal before the rf-SQUIDs, and faster sensors can be read out with lower-$Q$ resonators that allow for higher sampling rates above 40 kHz.

In summary, we have shown that microwave SQUID multiplexing readout is an excellent candidate for future large focal-plane arrays of spectroscopic sensors. In contrast to previous microwave measurements of cryogenic X-ray/gamma-ray sensors  \cite{Mazin_X-ray_MKIDs_APL_2006, Moore_PhononMKID_APL_2012}, we have demonstrated energy resolutions substantially better than conventional semiconducting detectors and that approach state-of-the-art results using traditional low-frequency readout. Dramatic increases in pixel count per readout channel will be straightforward to achieve by coupling additional resonators to the same feedline in order to use more of the 10 GHz of available HEMT bandwidth. Potential electromagnetic crosstalk between resonators in large arrays can be avoided by properly designing the resonators and embedding circuitry \cite{Noroozian_Crosstalk_IEEEMTT_2012, Noroozian_Thesis}. In the near term, the multiplexing factor will be constrained by the availability of multichannel microwave electronics to synthesize and demodulate large numbers of readout tones. This type of signal processing has recently been demonstrated for 256 sensors using high-performance but low-cost commercial electronics \cite{Mchugh_Readoutelecs_RSI_2012} and further improvements are certain. Our approach is compatible with TES sensors designed for other applications such as low-energy X-ray spectroscopy and the detection of single optical photons \cite{Cabrera_OpticalTES_APL_1998, Gerrits_OpticalTES_2012} as well as with magnetic calorimeters \cite{Bandler_MagneticMicrocalorimeters_JLTP_2012}. Even larger increases in the multiplexing factor will be achievable by embedding code-division multiplexed sensor columns in each microwave resonator \cite{Irwin_CDM_JLTP_2012}.

This work was supported by DHS under grant 2011-DN-077-ARI051, by the US Department
of Energy through the Office of Nonproliferation Research and Development and the Office of Nuclear Energy, and by NASA, under contract NNH11AR83I. The authors thank A. Betz and C. Bockstiegel for useful discussions and help with the experiment.


%
%

%



\bibliography{bibfile_arXiv}

\begin{thebibliography}{27}%
\makeatletter
\providecommand \@ifxundefined [1]{%
 \@ifx{#1\undefined}
}%
\providecommand \@ifnum [1]{%
 \ifnum #1\expandafter \@firstoftwo
 \else \expandafter \@secondoftwo
 \fi
}%
\providecommand \@ifx [1]{%
 \ifx #1\expandafter \@firstoftwo
 \else \expandafter \@secondoftwo
 \fi
}%
\providecommand \natexlab [1]{#1}%
\providecommand \enquote  [1]{``#1''}%
\providecommand \bibnamefont  [1]{#1}%
\providecommand \bibfnamefont [1]{#1}%
\providecommand \citenamefont [1]{#1}%
\providecommand \href@noop [0]{\@secondoftwo}%
\providecommand \href [0]{\begingroup \@sanitize@url \@href}%
\providecommand \@href[1]{\@@startlink{#1}\@@href}%
\providecommand \@@href[1]{\endgroup#1\@@endlink}%
\providecommand \@sanitize@url [0]{\catcode `\\12\catcode `\$12\catcode
  `\&12\catcode `\#12\catcode `\^12\catcode `\_12\catcode `\%12\relax}%
\providecommand \@@startlink[1]{}%
\providecommand \@@endlink[0]{}%
\providecommand \url  [0]{\begingroup\@sanitize@url \@url }%
\providecommand \@url [1]{\endgroup\@href {#1}{\urlprefix }}%
\providecommand \urlprefix  [0]{URL }%
\providecommand \Eprint [0]{\href }%
\providecommand \doibase [0]{http://dx.doi.org/}%
\providecommand \selectlanguage [0]{\@gobble}%
\providecommand \bibinfo  [0]{\@secondoftwo}%
\providecommand \bibfield  [0]{\@secondoftwo}%
\providecommand \translation [1]{[#1]}%
\providecommand \BibitemOpen [0]{}%
\providecommand \bibitemStop [0]{}%
\providecommand \bibitemNoStop [0]{.\EOS\space}%
\providecommand \EOS [0]{\spacefactor3000\relax}%
\providecommand \BibitemShut  [1]{\csname bibitem#1\endcsname}%
\let\auto@bib@innerbib\@empty
\bibitem [{\citenamefont {Doriese}\ \emph {et~al.}(2004)\citenamefont
  {Doriese}, \citenamefont {Beall}, \citenamefont {Deiker}, \citenamefont
  {Duncan}, \citenamefont {Ferreira}, \citenamefont {Hilton}, \citenamefont
  {Irwin}, \citenamefont {Reintsema}, \citenamefont {Ullom}, \citenamefont
  {Vale},\ and\ \citenamefont {Xu}}]{Doriese_TDM_APL_2004}%
  \BibitemOpen
  \bibfield  {author} {\bibinfo {author} {\bibfnamefont {W.~B.}\ \bibnamefont
  {Doriese}}, \bibinfo {author} {\bibfnamefont {J.~A.}\ \bibnamefont {Beall}},
  \bibinfo {author} {\bibfnamefont {S.}~\bibnamefont {Deiker}}, \bibinfo
  {author} {\bibfnamefont {W.~D.}\ \bibnamefont {Duncan}}, \bibinfo {author}
  {\bibfnamefont {L.}~\bibnamefont {Ferreira}}, \bibinfo {author}
  {\bibfnamefont {G.~C.}\ \bibnamefont {Hilton}}, \bibinfo {author}
  {\bibfnamefont {K.~D.}\ \bibnamefont {Irwin}}, \bibinfo {author}
  {\bibfnamefont {C.~D.}\ \bibnamefont {Reintsema}}, \bibinfo {author}
  {\bibfnamefont {J.~N.}\ \bibnamefont {Ullom}}, \bibinfo {author}
  {\bibfnamefont {L.~R.}\ \bibnamefont {Vale}}, \ and\ \bibinfo {author}
  {\bibfnamefont {Y.}~\bibnamefont {Xu}},\ }\href {\doibase 10.1063/1.1823041}
  {\bibfield  {journal} {\bibinfo  {journal} {Appl. Phys. Lett.}\ }\textbf
  {\bibinfo {volume} {85}},\ \bibinfo {pages} {4762--4764} (\bibinfo {year}
  {2004})}\BibitemShut {NoStop}%
\bibitem [{\citenamefont {Cunningham}\ \emph {et~al.}(2002)\citenamefont
  {Cunningham}, \citenamefont {Ullom}, \citenamefont {Miyazaki}, \citenamefont
  {Labov}, \citenamefont {Clarke}, \citenamefont {Lanting}, \citenamefont
  {Lee}, \citenamefont {Richards}, \citenamefont {Yoon},\ and\ \citenamefont
  {Spieler}}]{Cunningham_FDM_APL_2002}%
  \BibitemOpen
  \bibfield  {author} {\bibinfo {author} {\bibfnamefont {M.~F.}\ \bibnamefont
  {Cunningham}}, \bibinfo {author} {\bibfnamefont {J.~N.}\ \bibnamefont
  {Ullom}}, \bibinfo {author} {\bibfnamefont {T.}~\bibnamefont {Miyazaki}},
  \bibinfo {author} {\bibfnamefont {S.~E.}\ \bibnamefont {Labov}}, \bibinfo
  {author} {\bibfnamefont {J.}~\bibnamefont {Clarke}}, \bibinfo {author}
  {\bibfnamefont {T.~M.}\ \bibnamefont {Lanting}}, \bibinfo {author}
  {\bibfnamefont {A.~T.}\ \bibnamefont {Lee}}, \bibinfo {author} {\bibfnamefont
  {P.~L.}\ \bibnamefont {Richards}}, \bibinfo {author} {\bibfnamefont
  {J.}~\bibnamefont {Yoon}}, \ and\ \bibinfo {author} {\bibfnamefont
  {H.}~\bibnamefont {Spieler}},\ }\href {\doibase 10.1063/1.1489486} {\bibfield
   {journal} {\bibinfo  {journal} {Appl. Phys. Lett.}\ }\textbf {\bibinfo
  {volume} {81}},\ \bibinfo {pages} {159--161} (\bibinfo {year}
  {2002})}\BibitemShut {NoStop}%
\bibitem [{\citenamefont {Stiehl}\ \emph {et~al.}(2012)\citenamefont {Stiehl},
  \citenamefont {Doriese}, \citenamefont {Fowler}, \citenamefont {Hilton},
  \citenamefont {Irwin}, \citenamefont {Reintsema}, \citenamefont {Schmidt},
  \citenamefont {Swetz}, \citenamefont {Ullom},\ and\ \citenamefont
  {Vale}}]{Stiehl_CDM_APL_2012}%
  \BibitemOpen
  \bibfield  {author} {\bibinfo {author} {\bibfnamefont {G.~M.}\ \bibnamefont
  {Stiehl}}, \bibinfo {author} {\bibfnamefont {W.~B.}\ \bibnamefont {Doriese}},
  \bibinfo {author} {\bibfnamefont {J.~W.}\ \bibnamefont {Fowler}}, \bibinfo
  {author} {\bibfnamefont {G.~C.}\ \bibnamefont {Hilton}}, \bibinfo {author}
  {\bibfnamefont {K.~D.}\ \bibnamefont {Irwin}}, \bibinfo {author}
  {\bibfnamefont {C.~D.}\ \bibnamefont {Reintsema}}, \bibinfo {author}
  {\bibfnamefont {D.~R.}\ \bibnamefont {Schmidt}}, \bibinfo {author}
  {\bibfnamefont {D.~S.}\ \bibnamefont {Swetz}}, \bibinfo {author}
  {\bibfnamefont {J.~N.}\ \bibnamefont {Ullom}}, \ and\ \bibinfo {author}
  {\bibfnamefont {L.~R.}\ \bibnamefont {Vale}},\ }\href {\doibase
  10.1063/1.3684807} {\bibfield  {journal} {\bibinfo  {journal} {Appl. Phys.
  Lett.}\ }\textbf {\bibinfo {volume} {100}},\ \bibinfo {eid} {072601}
  (\bibinfo {year} {2012})}\BibitemShut {NoStop}%
\bibitem [{\citenamefont {Holland}\ \emph {et~al.}(2013)\citenamefont
  {Holland}, \citenamefont {Bintley}, \citenamefont {Chapin}, \citenamefont
  {Chrysostomou}, \citenamefont {Davis}, \citenamefont {Dempsey}, \citenamefont
  {Duncan}, \citenamefont {Fich}, \citenamefont {Friberg}, \citenamefont
  {Halpern}, \citenamefont {Irwin}, \citenamefont {Jenness}, \citenamefont
  {Kelly}, \citenamefont {MacIntosh}, \citenamefont {Robson}, \citenamefont
  {Scott}, \citenamefont {Ade}, \citenamefont {Atad-Ettedgui}, \citenamefont
  {Berry}, \citenamefont {Craig}, \citenamefont {Gao}, \citenamefont {Gibb},
  \citenamefont {Hilton}, \citenamefont {Hollister}, \citenamefont {Kycia},
  \citenamefont {Lunney}, \citenamefont {McGregor}, \citenamefont {Montgomery},
  \citenamefont {Parkes}, \citenamefont {Tilanus}, \citenamefont {Ullom},
  \citenamefont {Walther}, \citenamefont {Walton}, \citenamefont {Woodcraft},
  \citenamefont {Amiri}, \citenamefont {Atkinson}, \citenamefont {Burger},
  \citenamefont {Chuter}, \citenamefont {Coulson}, \citenamefont {Doriese},
  \citenamefont {Dunare}, \citenamefont {Economou}, \citenamefont {Niemack},
  \citenamefont {Parsons}, \citenamefont {Reintsema}, \citenamefont
  {Sibthorpe}, \citenamefont {Smail}, \citenamefont {Sudiwala},\ and\
  \citenamefont {Thomas}}]{Holland_SCUBA-2_MNRAS_2013}%
  \BibitemOpen
  \bibfield  {author} {\bibinfo {author} {\bibfnamefont {W.~S.}\ \bibnamefont
  {Holland}}, \bibinfo {author} {\bibfnamefont {D.}~\bibnamefont {Bintley}},
  \bibinfo {author} {\bibfnamefont {E.~L.}\ \bibnamefont {Chapin}}, \bibinfo
  {author} {\bibfnamefont {A.}~\bibnamefont {Chrysostomou}}, \bibinfo {author}
  {\bibfnamefont {G.~R.}\ \bibnamefont {Davis}}, \bibinfo {author}
  {\bibfnamefont {J.~T.}\ \bibnamefont {Dempsey}}, \bibinfo {author}
  {\bibfnamefont {W.~D.}\ \bibnamefont {Duncan}}, \bibinfo {author}
  {\bibfnamefont {M.}~\bibnamefont {Fich}}, \bibinfo {author} {\bibfnamefont
  {P.}~\bibnamefont {Friberg}}, \bibinfo {author} {\bibfnamefont
  {M.}~\bibnamefont {Halpern}}, \bibinfo {author} {\bibfnamefont {K.~D.}\
  \bibnamefont {Irwin}}, \bibinfo {author} {\bibfnamefont {T.}~\bibnamefont
  {Jenness}}, \bibinfo {author} {\bibfnamefont {B.~D.}\ \bibnamefont {Kelly}},
  \bibinfo {author} {\bibfnamefont {M.~J.}\ \bibnamefont {MacIntosh}}, \bibinfo
  {author} {\bibfnamefont {E.~I.}\ \bibnamefont {Robson}}, \bibinfo {author}
  {\bibfnamefont {D.}~\bibnamefont {Scott}}, \bibinfo {author} {\bibfnamefont
  {P.~A.~R.}\ \bibnamefont {Ade}}, \bibinfo {author} {\bibfnamefont
  {E.}~\bibnamefont {Atad-Ettedgui}}, \bibinfo {author} {\bibfnamefont {D.~S.}\
  \bibnamefont {Berry}}, \bibinfo {author} {\bibfnamefont {S.~C.}\ \bibnamefont
  {Craig}}, \bibinfo {author} {\bibfnamefont {X.}~\bibnamefont {Gao}}, \bibinfo
  {author} {\bibfnamefont {A.~G.}\ \bibnamefont {Gibb}}, \bibinfo {author}
  {\bibfnamefont {G.~C.}\ \bibnamefont {Hilton}}, \bibinfo {author}
  {\bibfnamefont {M.~I.}\ \bibnamefont {Hollister}}, \bibinfo {author}
  {\bibfnamefont {J.~B.}\ \bibnamefont {Kycia}}, \bibinfo {author}
  {\bibfnamefont {D.~W.}\ \bibnamefont {Lunney}}, \bibinfo {author}
  {\bibfnamefont {H.}~\bibnamefont {McGregor}}, \bibinfo {author}
  {\bibfnamefont {D.}~\bibnamefont {Montgomery}}, \bibinfo {author}
  {\bibfnamefont {W.}~\bibnamefont {Parkes}}, \bibinfo {author} {\bibfnamefont
  {R.~P.~J.}\ \bibnamefont {Tilanus}}, \bibinfo {author} {\bibfnamefont
  {J.~N.}\ \bibnamefont {Ullom}}, \bibinfo {author} {\bibfnamefont {C.~A.}\
  \bibnamefont {Walther}}, \bibinfo {author} {\bibfnamefont {A.~J.}\
  \bibnamefont {Walton}}, \bibinfo {author} {\bibfnamefont {A.~L.}\
  \bibnamefont {Woodcraft}}, \bibinfo {author} {\bibfnamefont {M.}~\bibnamefont
  {Amiri}}, \bibinfo {author} {\bibfnamefont {D.}~\bibnamefont {Atkinson}},
  \bibinfo {author} {\bibfnamefont {B.}~\bibnamefont {Burger}}, \bibinfo
  {author} {\bibfnamefont {T.}~\bibnamefont {Chuter}}, \bibinfo {author}
  {\bibfnamefont {I.~M.}\ \bibnamefont {Coulson}}, \bibinfo {author}
  {\bibfnamefont {W.~B.}\ \bibnamefont {Doriese}}, \bibinfo {author}
  {\bibfnamefont {C.}~\bibnamefont {Dunare}}, \bibinfo {author} {\bibfnamefont
  {F.}~\bibnamefont {Economou}}, \bibinfo {author} {\bibfnamefont {M.~D.}\
  \bibnamefont {Niemack}}, \bibinfo {author} {\bibfnamefont {H.~A.~L.}\
  \bibnamefont {Parsons}}, \bibinfo {author} {\bibfnamefont {C.~D.}\
  \bibnamefont {Reintsema}}, \bibinfo {author} {\bibfnamefont {B.}~\bibnamefont
  {Sibthorpe}}, \bibinfo {author} {\bibfnamefont {I.}~\bibnamefont {Smail}},
  \bibinfo {author} {\bibfnamefont {R.}~\bibnamefont {Sudiwala}}, \ and\
  \bibinfo {author} {\bibfnamefont {H.~S.}\ \bibnamefont {Thomas}},\ }\href
  {\doibase 10.1093/mnras/sts612} {\bibfield  {journal} {\bibinfo  {journal}
  {Monthly Notices of the Royal Astronomical Society}\ }\textbf {\bibinfo
  {volume} {430}},\ \bibinfo {pages} {2513--2533} (\bibinfo {year}
  {2013})}\BibitemShut {NoStop}%
\bibitem [{\citenamefont {Bennett}\ \emph {et~al.}(2012)\citenamefont
  {Bennett}, \citenamefont {Horansky}, \citenamefont {Schmidt}, \citenamefont
  {Hoover}, \citenamefont {Winkler}, \citenamefont {Alpert}, \citenamefont
  {Beall}, \citenamefont {Doriese}, \citenamefont {Fowler}, \citenamefont
  {Fitzgerald}, \citenamefont {Hilton}, \citenamefont {Irwin}, \citenamefont
  {Kotsubo}, \citenamefont {Mates}, \citenamefont {O'Neil}, \citenamefont
  {Rabin}, \citenamefont {Reintsema}, \citenamefont {Schima}, \citenamefont
  {Swetz}, \citenamefont {Vale},\ and\ \citenamefont
  {Ullom}}]{Bennett_microcal_SciRev_2012}%
  \BibitemOpen
  \bibfield  {author} {\bibinfo {author} {\bibfnamefont {D.~A.}\ \bibnamefont
  {Bennett}}, \bibinfo {author} {\bibfnamefont {R.~D.}\ \bibnamefont
  {Horansky}}, \bibinfo {author} {\bibfnamefont {D.~R.}\ \bibnamefont
  {Schmidt}}, \bibinfo {author} {\bibfnamefont {A.~S.}\ \bibnamefont {Hoover}},
  \bibinfo {author} {\bibfnamefont {R.}~\bibnamefont {Winkler}}, \bibinfo
  {author} {\bibfnamefont {B.~K.}\ \bibnamefont {Alpert}}, \bibinfo {author}
  {\bibfnamefont {J.~A.}\ \bibnamefont {Beall}}, \bibinfo {author}
  {\bibfnamefont {W.~B.}\ \bibnamefont {Doriese}}, \bibinfo {author}
  {\bibfnamefont {J.~W.}\ \bibnamefont {Fowler}}, \bibinfo {author}
  {\bibfnamefont {C.~P.}\ \bibnamefont {Fitzgerald}}, \bibinfo {author}
  {\bibfnamefont {G.~C.}\ \bibnamefont {Hilton}}, \bibinfo {author}
  {\bibfnamefont {K.~D.}\ \bibnamefont {Irwin}}, \bibinfo {author}
  {\bibfnamefont {V.}~\bibnamefont {Kotsubo}}, \bibinfo {author} {\bibfnamefont
  {J.~A.~B.}\ \bibnamefont {Mates}}, \bibinfo {author} {\bibfnamefont {G.~C.}\
  \bibnamefont {O'Neil}}, \bibinfo {author} {\bibfnamefont {M.~W.}\
  \bibnamefont {Rabin}}, \bibinfo {author} {\bibfnamefont {C.~D.}\ \bibnamefont
  {Reintsema}}, \bibinfo {author} {\bibfnamefont {F.~J.}\ \bibnamefont
  {Schima}}, \bibinfo {author} {\bibfnamefont {D.~S.}\ \bibnamefont {Swetz}},
  \bibinfo {author} {\bibfnamefont {L.~R.}\ \bibnamefont {Vale}}, \ and\
  \bibinfo {author} {\bibfnamefont {J.~N.}\ \bibnamefont {Ullom}},\ }\href
  {\doibase 10.1063/1.4754630} {\bibfield  {journal} {\bibinfo  {journal} {Rev.
  Sci. Instrum.}\ }\textbf {\bibinfo {volume} {83}},\ \bibinfo {eid} {093113}
  (\bibinfo {year} {2012})}\BibitemShut {NoStop}%
\bibitem [{\citenamefont {Day}\ \emph {et~al.}(2003)\citenamefont {Day},
  \citenamefont {LeDuc}, \citenamefont {Mazin}, \citenamefont {Vayonakis},\
  and\ \citenamefont {Zmuidzinas}}]{Day_MKIDs_Nature_2003}%
  \BibitemOpen
  \bibfield  {author} {\bibinfo {author} {\bibfnamefont {P.}~\bibnamefont
  {Day}}, \bibinfo {author} {\bibfnamefont {H.}~\bibnamefont {LeDuc}}, \bibinfo
  {author} {\bibfnamefont {B.}~\bibnamefont {Mazin}}, \bibinfo {author}
  {\bibfnamefont {A.}~\bibnamefont {Vayonakis}}, \ and\ \bibinfo {author}
  {\bibfnamefont {J.}~\bibnamefont {Zmuidzinas}},\ }\href {\doibase
  10.1038/nature02037} {\bibfield  {journal} {\bibinfo  {journal} {Nature}\
  }\textbf {\bibinfo {volume} {425}},\ \bibinfo {pages} {817--821} (\bibinfo
  {year} {2003})}\BibitemShut {NoStop}%
\bibitem [{\citenamefont
  {Zmuidzinas}(2012)}]{Zmuidzinas_ResonatorsReview_ARCMP_2012}%
  \BibitemOpen
  \bibfield  {author} {\bibinfo {author} {\bibfnamefont {J.}~\bibnamefont
  {Zmuidzinas}},\ }\href {\doibase 10.1146/annurev-conmatphys-020911-125022}
  {\bibfield  {journal} {\bibinfo  {journal} {Annu. Rev. Cond. Mater. Phys.}\
  }\textbf {\bibinfo {volume} {3}},\ \bibinfo {pages} {169--214} (\bibinfo
  {year} {2012})}\BibitemShut {NoStop}%
\bibitem [{\citenamefont {McHugh}\ \emph {et~al.}(2012)\citenamefont {McHugh},
  \citenamefont {Mazin}, \citenamefont {Serfass}, \citenamefont {Meeker},
  \citenamefont {O'Brien}, \citenamefont {Duan}, \citenamefont {Raffanti},\
  and\ \citenamefont {Werthimer}}]{Mchugh_Readoutelecs_RSI_2012}%
  \BibitemOpen
  \bibfield  {author} {\bibinfo {author} {\bibfnamefont {S.}~\bibnamefont
  {McHugh}}, \bibinfo {author} {\bibfnamefont {B.~A.}\ \bibnamefont {Mazin}},
  \bibinfo {author} {\bibfnamefont {B.}~\bibnamefont {Serfass}}, \bibinfo
  {author} {\bibfnamefont {S.}~\bibnamefont {Meeker}}, \bibinfo {author}
  {\bibfnamefont {K.}~\bibnamefont {O'Brien}}, \bibinfo {author} {\bibfnamefont
  {R.}~\bibnamefont {Duan}}, \bibinfo {author} {\bibfnamefont {R.}~\bibnamefont
  {Raffanti}}, \ and\ \bibinfo {author} {\bibfnamefont {D.}~\bibnamefont
  {Werthimer}},\ }\href {\doibase 10.1063/1.3700812} {\bibfield  {journal}
  {\bibinfo  {journal} {Rev. Sci. Instrum.}\ }\textbf {\bibinfo {volume}
  {83}},\ \bibinfo {eid} {044702} (\bibinfo {year} {2012})}\BibitemShut
  {NoStop}%
\bibitem [{\citenamefont {Mazin}\ \emph {et~al.}(2006)\citenamefont {Mazin},
  \citenamefont {Bumble}, \citenamefont {Day}, \citenamefont {Eckart},
  \citenamefont {Golwala}, \citenamefont {Zmuidzinas},\ and\ \citenamefont
  {Harrison}}]{Mazin_X-ray_MKIDs_APL_2006}%
  \BibitemOpen
  \bibfield  {author} {\bibinfo {author} {\bibfnamefont {B.~A.}\ \bibnamefont
  {Mazin}}, \bibinfo {author} {\bibfnamefont {B.}~\bibnamefont {Bumble}},
  \bibinfo {author} {\bibfnamefont {P.~K.}\ \bibnamefont {Day}}, \bibinfo
  {author} {\bibfnamefont {M.~E.}\ \bibnamefont {Eckart}}, \bibinfo {author}
  {\bibfnamefont {S.}~\bibnamefont {Golwala}}, \bibinfo {author} {\bibfnamefont
  {J.}~\bibnamefont {Zmuidzinas}}, \ and\ \bibinfo {author} {\bibfnamefont
  {F.~A.}\ \bibnamefont {Harrison}},\ }\href {\doibase 10.1063/1.2390664}
  {\bibfield  {journal} {\bibinfo  {journal} {Appl. Phys. Lett.}\ }\textbf
  {\bibinfo {volume} {89}},\ \bibinfo {eid} {222507} (\bibinfo {year}
  {2006})}\BibitemShut {NoStop}%
\bibitem [{\citenamefont {Moore}\ \emph {et~al.}(2012)\citenamefont {Moore},
  \citenamefont {Golwala}, \citenamefont {Bumble}, \citenamefont {Cornell},
  \citenamefont {Day}, \citenamefont {LeDuc},\ and\ \citenamefont
  {Zmuidzinas}}]{Moore_PhononMKID_APL_2012}%
  \BibitemOpen
  \bibfield  {author} {\bibinfo {author} {\bibfnamefont {D.~C.}\ \bibnamefont
  {Moore}}, \bibinfo {author} {\bibfnamefont {S.~R.}\ \bibnamefont {Golwala}},
  \bibinfo {author} {\bibfnamefont {B.}~\bibnamefont {Bumble}}, \bibinfo
  {author} {\bibfnamefont {B.}~\bibnamefont {Cornell}}, \bibinfo {author}
  {\bibfnamefont {P.~K.}\ \bibnamefont {Day}}, \bibinfo {author} {\bibfnamefont
  {H.~G.}\ \bibnamefont {LeDuc}}, \ and\ \bibinfo {author} {\bibfnamefont
  {J.}~\bibnamefont {Zmuidzinas}},\ }\href {\doibase 10.1063/1.4726279}
  {\bibfield  {journal} {\bibinfo  {journal} {Appl. Phys. Lett.}\ }\textbf
  {\bibinfo {volume} {100}},\ \bibinfo {eid} {232601} (\bibinfo {year}
  {2012})}\BibitemShut {NoStop}%
\bibitem [{\citenamefont {Irwin}\ and\ \citenamefont
  {Lehnert}(2004)}]{Irwin_MicrowaveSQUIDMux_APL_2004}%
  \BibitemOpen
  \bibfield  {author} {\bibinfo {author} {\bibfnamefont {K.~D.}\ \bibnamefont
  {Irwin}}\ and\ \bibinfo {author} {\bibfnamefont {K.~W.}\ \bibnamefont
  {Lehnert}},\ }\href {\doibase 10.1063/1.1791733} {\bibfield  {journal}
  {\bibinfo  {journal} {Appl. Phys. Lett.}\ }\textbf {\bibinfo {volume} {85}},\
  \bibinfo {pages} {2107--2109} (\bibinfo {year} {2004})}\BibitemShut {NoStop}%
\bibitem [{\citenamefont {Mates}\ \emph {et~al.}(2008)\citenamefont {Mates},
  \citenamefont {Hilton}, \citenamefont {Irwin}, \citenamefont {Vale},\ and\
  \citenamefont {Lehnert}}]{Mates_MicrowaveSQUIDMux_APL_2008}%
  \BibitemOpen
  \bibfield  {author} {\bibinfo {author} {\bibfnamefont {J.~A.~B.}\
  \bibnamefont {Mates}}, \bibinfo {author} {\bibfnamefont {G.~C.}\ \bibnamefont
  {Hilton}}, \bibinfo {author} {\bibfnamefont {K.~D.}\ \bibnamefont {Irwin}},
  \bibinfo {author} {\bibfnamefont {L.~R.}\ \bibnamefont {Vale}}, \ and\
  \bibinfo {author} {\bibfnamefont {K.~W.}\ \bibnamefont {Lehnert}},\ }\href
  {\doibase 10.1063/1.2803852} {\bibfield  {journal} {\bibinfo  {journal}
  {Appl. Phys. Lett.}\ }\textbf {\bibinfo {volume} {92}},\ \bibinfo {eid}
  {023514} (\bibinfo {year} {2008})}\BibitemShut {NoStop}%
\bibitem [{\citenamefont {Mates}(2011)}]{Mates_Thesis}%
  \BibitemOpen
  \bibfield  {author} {\bibinfo {author} {\bibfnamefont {J.}~\bibnamefont
  {Mates}},\ }\emph {\bibinfo {title} {The Microwave SQUID Multiplexer}},\
  \href@noop {} {Ph.D. thesis},\ \bibinfo  {school} {University of Colorado,
  Boulder} (\bibinfo {year} {2011})\BibitemShut {NoStop}%
\bibitem [{\citenamefont {Mates}\ \emph {et~al.}(2012)\citenamefont {Mates},
  \citenamefont {Irwin}, \citenamefont {Vale}, \citenamefont {Hilton},
  \citenamefont {Gao},\ and\ \citenamefont
  {Lehnert}}]{Mates_FluxrampModulation_JLTP_2012}%
  \BibitemOpen
  \bibfield  {author} {\bibinfo {author} {\bibfnamefont {J.~A.~B.}\
  \bibnamefont {Mates}}, \bibinfo {author} {\bibfnamefont {K.~D.}\ \bibnamefont
  {Irwin}}, \bibinfo {author} {\bibfnamefont {L.~R.}\ \bibnamefont {Vale}},
  \bibinfo {author} {\bibfnamefont {G.~C.}\ \bibnamefont {Hilton}}, \bibinfo
  {author} {\bibfnamefont {J.}~\bibnamefont {Gao}}, \ and\ \bibinfo {author}
  {\bibfnamefont {K.~W.}\ \bibnamefont {Lehnert}},\ }\href {\doibase
  10.1007/s10909-012-0518-6} {\bibfield  {journal} {\bibinfo  {journal} {J. Low
  Temp. Phys.}\ }\textbf {\bibinfo {volume} {167}},\ \bibinfo {pages}
  {707--712} (\bibinfo {year} {2012})}\BibitemShut {NoStop}%
\bibitem [{\citenamefont {Noroozian}()}]{uMux_Movie}%
  \BibitemOpen
  \bibfield  {author} {\bibinfo {author} {\bibfnamefont {O.}~\bibnamefont
  {Noroozian}},\ }\href {http://youtu.be/I9AX5YSGgdU} {\enquote {\bibinfo
  {title} {http://youtu.be/i9ax5ysggdu},}\ }\BibitemShut {NoStop}%
\bibitem [{\citenamefont {Duan}\ \emph {et~al.}(2010)\citenamefont {Duan},
  \citenamefont {McHugh}, \citenamefont {Serfass}, \citenamefont {Mazin},
  \citenamefont {Merrill}, \citenamefont {Golwala}, \citenamefont {Downes},
  \citenamefont {Czakon}, \citenamefont {Day}, \citenamefont {Gao},
  \citenamefont {Glenn}, \citenamefont {Hollister}, \citenamefont {Leduc},
  \citenamefont {Maloney}, \citenamefont {Noroozian}, \citenamefont {Nguyen},
  \citenamefont {Sayers}, \citenamefont {Schlaerth}, \citenamefont {Siegel},
  \citenamefont {Vaillancourt}, \citenamefont {Vayonakis}, \citenamefont
  {Wilson},\ and\ \citenamefont
  {Zmuidzinas}}]{Duan_Opensource_readout_SPIE_2010}%
  \BibitemOpen
  \bibfield  {author} {\bibinfo {author} {\bibfnamefont {R.}~\bibnamefont
  {Duan}}, \bibinfo {author} {\bibfnamefont {S.}~\bibnamefont {McHugh}},
  \bibinfo {author} {\bibfnamefont {B.}~\bibnamefont {Serfass}}, \bibinfo
  {author} {\bibfnamefont {B.~A.}\ \bibnamefont {Mazin}}, \bibinfo {author}
  {\bibfnamefont {A.}~\bibnamefont {Merrill}}, \bibinfo {author} {\bibfnamefont
  {S.~R.}\ \bibnamefont {Golwala}}, \bibinfo {author} {\bibfnamefont {T.~P.}\
  \bibnamefont {Downes}}, \bibinfo {author} {\bibfnamefont {N.~G.}\
  \bibnamefont {Czakon}}, \bibinfo {author} {\bibfnamefont {P.~K.}\
  \bibnamefont {Day}}, \bibinfo {author} {\bibfnamefont {J.}~\bibnamefont
  {Gao}}, \bibinfo {author} {\bibfnamefont {J.}~\bibnamefont {Glenn}}, \bibinfo
  {author} {\bibfnamefont {M.~I.}\ \bibnamefont {Hollister}}, \bibinfo {author}
  {\bibfnamefont {H.~G.}\ \bibnamefont {Leduc}}, \bibinfo {author}
  {\bibfnamefont {P.~R.}\ \bibnamefont {Maloney}}, \bibinfo {author}
  {\bibfnamefont {O.}~\bibnamefont {Noroozian}}, \bibinfo {author}
  {\bibfnamefont {H.~T.}\ \bibnamefont {Nguyen}}, \bibinfo {author}
  {\bibfnamefont {J.}~\bibnamefont {Sayers}}, \bibinfo {author} {\bibfnamefont
  {J.~A.}\ \bibnamefont {Schlaerth}}, \bibinfo {author} {\bibfnamefont
  {S.}~\bibnamefont {Siegel}}, \bibinfo {author} {\bibfnamefont {J.~E.}\
  \bibnamefont {Vaillancourt}}, \bibinfo {author} {\bibfnamefont
  {A.}~\bibnamefont {Vayonakis}}, \bibinfo {author} {\bibfnamefont {P.~R.}\
  \bibnamefont {Wilson}}, \ and\ \bibinfo {author} {\bibfnamefont
  {J.}~\bibnamefont {Zmuidzinas}},\ }\href
  {http://dx.doi.org/10.1117/12.856832} {\bibfield  {journal} {\bibinfo
  {journal} {Proc. SPIE}\ }\textbf {\bibinfo {volume} {7741}},\ \bibinfo
  {pages} {77411V} (\bibinfo {year} {2010})}\BibitemShut {NoStop}%
\bibitem [{\citenamefont {Gao}\ \emph {et~al.}(2007)\citenamefont {Gao},
  \citenamefont {Zmuidzinas}, \citenamefont {Mazin}, \citenamefont {LeDuc},\
  and\ \citenamefont {Day}}]{Gao_NoisePropertiesOfResonators_APL_2007}%
  \BibitemOpen
  \bibfield  {author} {\bibinfo {author} {\bibfnamefont {J.}~\bibnamefont
  {Gao}}, \bibinfo {author} {\bibfnamefont {J.}~\bibnamefont {Zmuidzinas}},
  \bibinfo {author} {\bibfnamefont {B.~A.}\ \bibnamefont {Mazin}}, \bibinfo
  {author} {\bibfnamefont {H.~G.}\ \bibnamefont {LeDuc}}, \ and\ \bibinfo
  {author} {\bibfnamefont {P.~K.}\ \bibnamefont {Day}},\ }\href
  {http://dx.doi.org/10.1063/1.2711770} {\bibfield  {journal} {\bibinfo
  {journal} {Appl. Phys. Lett.}\ }\textbf {\bibinfo {volume} {90}},\ \bibinfo
  {eid} {102507} (\bibinfo {year} {2007})}\BibitemShut {NoStop}%
\bibitem [{\citenamefont {Barends}\ \emph {et~al.}(2008)\citenamefont
  {Barends}, \citenamefont {Hortensius}, \citenamefont {Zijlstra},
  \citenamefont {Baselmans}, \citenamefont {Yates}, \citenamefont {Gao},\ and\
  \citenamefont {Klapwijk}}]{Barends_dielectricNoiseNbTiN_APL_2008}%
  \BibitemOpen
  \bibfield  {author} {\bibinfo {author} {\bibfnamefont {R.}~\bibnamefont
  {Barends}}, \bibinfo {author} {\bibfnamefont {H.~L.}\ \bibnamefont
  {Hortensius}}, \bibinfo {author} {\bibfnamefont {T.}~\bibnamefont
  {Zijlstra}}, \bibinfo {author} {\bibfnamefont {J.~J.~A.}\ \bibnamefont
  {Baselmans}}, \bibinfo {author} {\bibfnamefont {S.~J.~C.}\ \bibnamefont
  {Yates}}, \bibinfo {author} {\bibfnamefont {J.~R.}\ \bibnamefont {Gao}}, \
  and\ \bibinfo {author} {\bibfnamefont {T.~M.}\ \bibnamefont {Klapwijk}},\
  }\href {\doibase 10.1063/1.2937837} {\bibfield  {journal} {\bibinfo
  {journal} {Appl. Phys. Lett.}\ }\textbf {\bibinfo {volume} {92}},\ \bibinfo
  {eid} {223502} (\bibinfo {year} {2008})}\BibitemShut {NoStop}%
\bibitem [{\citenamefont {Noroozian}\ \emph {et~al.}(2009)\citenamefont
  {Noroozian}, \citenamefont {Gao}, \citenamefont {Zmuidzinas}, \citenamefont
  {LeDuc},\ and\ \citenamefont {Mazin}}]{Noroozian_TLS_LTD13_2009}%
  \BibitemOpen
  \bibfield  {author} {\bibinfo {author} {\bibfnamefont {O.}~\bibnamefont
  {Noroozian}}, \bibinfo {author} {\bibfnamefont {J.}~\bibnamefont {Gao}},
  \bibinfo {author} {\bibfnamefont {J.}~\bibnamefont {Zmuidzinas}}, \bibinfo
  {author} {\bibfnamefont {H.~G.}\ \bibnamefont {LeDuc}}, \ and\ \bibinfo
  {author} {\bibfnamefont {B.~A.}\ \bibnamefont {Mazin}},\ }\href {\doibase
  10.1063/1.3292302} {\bibfield  {journal} {\bibinfo  {journal} {AIP Conf.
  Proc.}\ }\textbf {\bibinfo {volume} {1185}},\ \bibinfo {pages} {148--151}
  (\bibinfo {year} {2009})}\BibitemShut {NoStop}%
\bibitem [{\citenamefont {Szymkowiak}\ \emph {et~al.}(1993)\citenamefont
  {Szymkowiak}, \citenamefont {Kelley}, \citenamefont {Moseley},\ and\
  \citenamefont {Stahle}}]{Szymkowiak_OptimalFilter_JLTP_1993}%
  \BibitemOpen
  \bibfield  {author} {\bibinfo {author} {\bibfnamefont {A.~E.}\ \bibnamefont
  {Szymkowiak}}, \bibinfo {author} {\bibfnamefont {R.~L.}\ \bibnamefont
  {Kelley}}, \bibinfo {author} {\bibfnamefont {S.~H.}\ \bibnamefont {Moseley}},
  \ and\ \bibinfo {author} {\bibfnamefont {C.~K.}\ \bibnamefont {Stahle}},\
  }\href {\doibase 10.1007/BF00693433} {\bibfield  {journal} {\bibinfo
  {journal} {J. Low Temp. Phys.}\ }\textbf {\bibinfo {volume} {93}},\ \bibinfo
  {pages} {281--285} (\bibinfo {year} {1993})}\BibitemShut {NoStop}%
\bibitem [{\citenamefont {Moseley}, \citenamefont {Mather},\ and\ \citenamefont
  {McCammon}(1984)}]{Moseley_xrayThermalDetectors_JAP_1984}%
  \BibitemOpen
  \bibfield  {author} {\bibinfo {author} {\bibfnamefont {S.~H.}\ \bibnamefont
  {Moseley}}, \bibinfo {author} {\bibfnamefont {J.~C.}\ \bibnamefont {Mather}},
  \ and\ \bibinfo {author} {\bibfnamefont {D.}~\bibnamefont {McCammon}},\
  }\href {\doibase 10.1063/1.334129} {\bibfield  {journal} {\bibinfo  {journal}
  {J. Appl. Phys.}\ }\textbf {\bibinfo {volume} {56}},\ \bibinfo {pages}
  {1257--1262} (\bibinfo {year} {1984})}\BibitemShut {NoStop}%
\bibitem [{\citenamefont {Noroozian}\ \emph {et~al.}(2012)\citenamefont
  {Noroozian}, \citenamefont {Day}, \citenamefont {Eom}, \citenamefont
  {Leduc},\ and\ \citenamefont
  {Zmuidzinas}}]{Noroozian_Crosstalk_IEEEMTT_2012}%
  \BibitemOpen
  \bibfield  {author} {\bibinfo {author} {\bibfnamefont {O.}~\bibnamefont
  {Noroozian}}, \bibinfo {author} {\bibfnamefont {P.}~\bibnamefont {Day}},
  \bibinfo {author} {\bibfnamefont {B.~H.}\ \bibnamefont {Eom}}, \bibinfo
  {author} {\bibfnamefont {H.}~\bibnamefont {Leduc}}, \ and\ \bibinfo {author}
  {\bibfnamefont {J.}~\bibnamefont {Zmuidzinas}},\ }\href {\doibase
  10.1109/TMTT.2012.2187538} {\bibfield  {journal} {\bibinfo  {journal} {IEEE
  Trans. Microwave Theory Tech.}\ }\textbf {\bibinfo {volume} {60}},\ \bibinfo
  {pages} {1235--1243} (\bibinfo {year} {2012})}\BibitemShut {NoStop}%
\bibitem [{\citenamefont {Noroozian}(2012)}]{Noroozian_Thesis}%
  \BibitemOpen
  \bibfield  {author} {\bibinfo {author} {\bibfnamefont {O.}~\bibnamefont
  {Noroozian}},\ }\emph {\bibinfo {title} {Superconducting Microwave Resonator
  Arrays for Submillimeter/Far-Infrared Imaging}},\ \href@noop {} {Ph.D.
  thesis},\ \bibinfo  {school} {California Institute of Technology, Pasadena}
  (\bibinfo {year} {2012})\BibitemShut {NoStop}%
\bibitem [{\citenamefont {Cabrera}\ \emph {et~al.}(1998)\citenamefont
  {Cabrera}, \citenamefont {Clarke}, \citenamefont {Colling}, \citenamefont
  {Miller}, \citenamefont {Nam},\ and\ \citenamefont
  {Romani}}]{Cabrera_OpticalTES_APL_1998}%
  \BibitemOpen
  \bibfield  {author} {\bibinfo {author} {\bibfnamefont {B.}~\bibnamefont
  {Cabrera}}, \bibinfo {author} {\bibfnamefont {R.~M.}\ \bibnamefont {Clarke}},
  \bibinfo {author} {\bibfnamefont {P.}~\bibnamefont {Colling}}, \bibinfo
  {author} {\bibfnamefont {A.~J.}\ \bibnamefont {Miller}}, \bibinfo {author}
  {\bibfnamefont {S.}~\bibnamefont {Nam}}, \ and\ \bibinfo {author}
  {\bibfnamefont {R.~W.}\ \bibnamefont {Romani}},\ }\href {\doibase
  10.1063/1.121984} {\bibfield  {journal} {\bibinfo  {journal} {Appl. Phys.
  Lett.}\ }\textbf {\bibinfo {volume} {73}},\ \bibinfo {pages} {735--737}
  (\bibinfo {year} {1998})}\BibitemShut {NoStop}%
\bibitem [{\citenamefont {{Gerrits}}\ \emph {et~al.}(2012)\citenamefont
  {{Gerrits}}, \citenamefont {{Calkins}}, \citenamefont {{Tomlin}},
  \citenamefont {{Lita}}, \citenamefont {{Migdall}}, \citenamefont {{Mirin}},\
  and\ \citenamefont {{Nam}}}]{Gerrits_OpticalTES_2012}%
  \BibitemOpen
  \bibfield  {author} {\bibinfo {author} {\bibfnamefont {T.}~\bibnamefont
  {{Gerrits}}}, \bibinfo {author} {\bibfnamefont {B.}~\bibnamefont
  {{Calkins}}}, \bibinfo {author} {\bibfnamefont {N.}~\bibnamefont {{Tomlin}}},
  \bibinfo {author} {\bibfnamefont {A.~E.}\ \bibnamefont {{Lita}}}, \bibinfo
  {author} {\bibfnamefont {A.}~\bibnamefont {{Migdall}}}, \bibinfo {author}
  {\bibfnamefont {R.}~\bibnamefont {{Mirin}}}, \ and\ \bibinfo {author}
  {\bibfnamefont {S.~W.}\ \bibnamefont {{Nam}}},\ }\href {\doibase
  10.1364/OE.20.023798} {\bibfield  {journal} {\bibinfo  {journal} {Opt.
  Express}\ }\textbf {\bibinfo {volume} {20}},\ \bibinfo {pages} {23798}
  (\bibinfo {year} {2012})}\BibitemShut {NoStop}%
\bibitem [{\citenamefont {Bandler}\ \emph {et~al.}(2012)\citenamefont
  {Bandler}, \citenamefont {Irwin}, \citenamefont {Kelly}, \citenamefont
  {Nagler}, \citenamefont {Porst}, \citenamefont {Rotzinger}, \citenamefont
  {Sadleir}, \citenamefont {Seidel}, \citenamefont {Smith},\ and\ \citenamefont
  {Stevenson}}]{Bandler_MagneticMicrocalorimeters_JLTP_2012}%
  \BibitemOpen
  \bibfield  {author} {\bibinfo {author} {\bibfnamefont {S.~R.}\ \bibnamefont
  {Bandler}}, \bibinfo {author} {\bibfnamefont {K.~D.}\ \bibnamefont {Irwin}},
  \bibinfo {author} {\bibfnamefont {D.}~\bibnamefont {Kelly}}, \bibinfo
  {author} {\bibfnamefont {P.~N.}\ \bibnamefont {Nagler}}, \bibinfo {author}
  {\bibfnamefont {J.-P.}\ \bibnamefont {Porst}}, \bibinfo {author}
  {\bibfnamefont {H.}~\bibnamefont {Rotzinger}}, \bibinfo {author}
  {\bibfnamefont {J.~E.}\ \bibnamefont {Sadleir}}, \bibinfo {author}
  {\bibfnamefont {G.~M.}\ \bibnamefont {Seidel}}, \bibinfo {author}
  {\bibfnamefont {S.~J.}\ \bibnamefont {Smith}}, \ and\ \bibinfo {author}
  {\bibfnamefont {T.~R.}\ \bibnamefont {Stevenson}},\ }\href {\doibase
  10.1007/s10909-012-0544-4} {\bibfield  {journal} {\bibinfo  {journal} {J. Low
  Temp. Phys.}\ }\textbf {\bibinfo {volume} {167}},\ \bibinfo {pages}
  {254--268} (\bibinfo {year} {2012})}\BibitemShut {NoStop}%
\bibitem [{\citenamefont {Irwin}\ \emph {et~al.}(2012)\citenamefont {Irwin},
  \citenamefont {Cho}, \citenamefont {Doriese}, \citenamefont {Fowler},
  \citenamefont {Hilton}, \citenamefont {Niemack}, \citenamefont {Reintsema},
  \citenamefont {Schmidt}, \citenamefont {Ullom},\ and\ \citenamefont
  {Vale}}]{Irwin_CDM_JLTP_2012}%
  \BibitemOpen
  \bibfield  {author} {\bibinfo {author} {\bibfnamefont {K.~D.}\ \bibnamefont
  {Irwin}}, \bibinfo {author} {\bibfnamefont {H.~M.}\ \bibnamefont {Cho}},
  \bibinfo {author} {\bibfnamefont {W.~B.}\ \bibnamefont {Doriese}}, \bibinfo
  {author} {\bibfnamefont {J.~W.}\ \bibnamefont {Fowler}}, \bibinfo {author}
  {\bibfnamefont {G.~C.}\ \bibnamefont {Hilton}}, \bibinfo {author}
  {\bibfnamefont {M.~D.}\ \bibnamefont {Niemack}}, \bibinfo {author}
  {\bibfnamefont {C.~D.}\ \bibnamefont {Reintsema}}, \bibinfo {author}
  {\bibfnamefont {D.~R.}\ \bibnamefont {Schmidt}}, \bibinfo {author}
  {\bibfnamefont {J.~N.}\ \bibnamefont {Ullom}}, \ and\ \bibinfo {author}
  {\bibfnamefont {L.~R.}\ \bibnamefont {Vale}},\ }\href {\doibase
  10.1007/s10909-012-0586-7} {\bibfield  {journal} {\bibinfo  {journal} {J. Low
  Temp. Phys.}\ }\textbf {\bibinfo {volume} {167}},\ \bibinfo {pages}
  {588--594} (\bibinfo {year} {2012})}\BibitemShut {NoStop}%
\end{thebibliography}%

\end{document}